# Using the Deutsch-Jozsa algorithm to determine parts of an array and apply a specified function to each independent part

Samir Lipovaca
slipovaca@nyse.com

*Abstract:* Using the Deutsch-Jozsa algorithm, we will develop a method for solving a class of problems in which we need to determine parts of an array and then apply a specified function to each independent part. Since present quantum computers are not robust enough for code writing and execution, we will build a model of a vector quantum computer that implements the Deutsch-Jozsa algorithm from a machine language view using the APL2 programming language. The core of the method is an operator (DJBOX) which allows evaluation of an arbitrary function $f$ by the Deutsch-Jozsa algorithm. Two key functions of the method are GET_PARTITION and CALC_WITH_PARTITIONS. The GET_PARTITION function determines parts of an array based on the function $f$. The CALC_WITH_PARTITIONS function determines parts of an array based on the function $f$ and then applies another function to each independent part. We will imagine the method is implemented on the above vector quantum computer. We will show that the method can be successfully executed.

**(1) Introduction**

The Deutsch-Jozsa algorithm is one of the first examples of a quantum algorithm. Quantum algorithms are designed for execution on quantum computers. Quantum algorithms have the potential to be more efficient than classical algorithms by taking advantage of the quantum superposition and entanglement principles. We are given a black box which takes $n$ qubits $x_1, x_2, \ldots, x_n$ and computes a $\{0,1\}$ valued function $f(x_1, x_2, \ldots, x_n)$. We know that the function is either constant for all inputs or else $f$ is balanced, that is, equal to 1 for exactly half of all possible inputs. The task is to determine whether $f$ is constant or balanced. The Deutsch-Jozsa algorithm produces an answer that is always correct with just 1 evaluation of $f$. For comparison, the best deterministic classical algorithm requires $\dfrac{2^n}{2} + 1$ queries.

A data structure is an arrangement of data in a computer's memory. Data structures include arrays, linked lists, stacks, binary trees, and hash tables, among others. Algorithms manipulate the data in these structures in various ways, such as searching for a particular data item and sorting the data. The array is the most commonly used data storage structure. It is built into most programming languages. Consecutive memory cells are used to store an array's data. For example, 10 consecutive memory cells store data for an array of 10 elements, one element per cell. Array elements are accessed using an index number. Usually, the first element is numbered 0, so that the indices in an array of 10 elements run from 0 to 9. For example, $A[4]$ returns content of fifth element of the array $A$.

Occasionally, it becomes necessary to apply a function independently to successive parts of an array. For example, the problem may be to multiply all elements in each part of an array $A$. If the array $A$ contains integers smaller than 10 and the parts were as follows:

$$2\ 2\ 3\ 1\ \ 1\ 2\ \ 7\ \ 5\ 2\ 1$$

then the desired result would be $12\ \ 2\ \ 7\ \ 10$. Essentially, in this class of problems [1] we need to determine parts (partitioning) of an array and then apply a specified function to each independent part. The function may be inherently difficult to calculate.

The following sections discuss an approach for solving this class of problems using the Deutsch-Jozsa algorithm. First, we must characterize the concept of "parts" and then combine that with the function and the array to produce the result. In the section 2.2 we describe partitioning in the APL2. The Deutsch-Jozsa algorithm is briefly outlined in the section 2.3. Then in the section 2.4 the APL2 simulation of the Deutsch-Jozsa algorithm is described. The APL2 simulation of a vector quantum computer that implements the Deutsch-Jozsa algorithm is exposed in the section 2.5. Finally, in the section 2.6, partitioning in the vector quantum computer that implements the Deutsch-Jozsa algorithm is described. We finish with Discussion and Conclusions.

**(2) Methods and Results**

*2.1 Partition array*

A simple and unique characterization [1] of any partitioning of an array $A$ is another array $B$ of the same length whose elements are 0 or 1. Let's call the array $B$ a partition array. Each 1 in the partition array corresponds to the beginning of a part and the following 0's correspond to the remaining elements in that part. The partition array used above would be characterized as

$$2\ 2\ 3\ 1\ \ 1\ 2\ \ 7\ \ 5\ 2\ 1$$
$$1\ 0\ 0\ 0\ \ 1\ 0\ \ 1\ \ 1\ 0\ 0$$

Any array of the same length as $A$ and whose elements are 0 or 1 and whose first element is 1 represents a partition array. It is easy to see that there are $2^{N-1}$ such partition arrays of length $N$. The number of parts represented by the partition array $B$ is the number of 1's in $B$. In a given problem, $B$ is defined or determined by some function.

*2.2 Partitioning in APL2*

APL2 is a computer programming language that permits rapid development of applications and models of application design. APL2 deals with whole collections of data all at once. Arrays are a fundamental unit of computation. In APL2, an array is a rectangular arrangement of data called the items of the array. Each item is a number, character or another array. Every row of an array contains the same number of items and every column in an array contains the same number of items. This rectangular design is extended to collections of data organized along any number of independent directions.

The directions along which data in an array is arranged are the axes of the array. The number of directions is the rank of the array. In APL2, ranks higher than 2 are allowed up to some implementation limit (typically 64). APL2 has special names for arrays that have rank 0, 1, or 2. A scalar is a single number (such as 5) or character that is not arranged along any axes and so has rank 0. A vector (such as 3 1 8 0) has one axis and a rank of 1. Notice in this terminology the array A in the section (1) is a vector. A matrix having rows and columns has two axes and a rank of 2.

Let's suppose the array $A$ has a rank of 2 and the following items

```
A  100
A  100
A   20
A  400
B   30
B  200
B  300
C  100
C  100
C  100
```

The problem is to accumulate items of the second column into bins $A, B$ and $C$. For example, we could imagine a large array A in which the first column contains individual atoms in a protein and the second column contains their respective interaction energies with the remaining atoms. Then the problem would be to calculate for each atom a total interaction energy with the rest of the protein. In this example the array $A$ is already ordered by the first column such that equal items are adjacent. Generally, the array $A$ must be sorted by respective column before the array $B$ is determined. The problem is solved in two steps. First, the array $B$ is determined by

$$B \leftarrow A[;1] \neq {}^{-}1 \phi A[;1]$$

where 1 refers to the first column of $A$. The effect of `¯1⌽A[;1]` is to move the last item of the first column of $A$ to be the first row item. `¯1⌽A[;1]` is displayed below in the second row. It is compared with `A[;1]` and whenever items are different, the value of 1 is assigned to $B$ (the third row below).

```
A  A  A   A   B  B   B   C  C  C
C  A  A   A   A  B   B   B  C  C
1  0  0   0   1  0   0   1  0  0
```

Then we calculate totals for each partition

$$T \leftarrow +/\ddot{}(+\backslash B) \subset A[;2]$$

where 2 refers to the second column of $A$. `(+\B)⊂A[;2]` divides up the second column of $A$ producing 3 parts and sum is applied to each part which is represented by `+/¨`. Combining all together a final array $R$

$$R \leftarrow (,[''] B/A[;1]), T$$

has the following elements

```
A  620
B  530
C  300
```

*2.3 The Deutsch-Jozsa algorithm*

Let us briefly outline specific steps of the algorithm [2]. The input state

$$|\psi_0\rangle = |0\rangle^{\otimes n}|1\rangle \qquad (1)$$

consists of the query register which describes the state of $n$ qubits all prepared in the $|0\rangle$ state and the answer register in the $|1\rangle$ state. After the Hadamard transform on the query register and the Hadamard gate on the answer register we obtain

$$|\psi_1\rangle = \sum_{x \in \{0,1\}^n} \frac{|x\rangle}{\sqrt{2^n}} \left(\frac{|0\rangle - |1\rangle}{\sqrt{2}}\right). \qquad (2)$$

The query register is a superposition of all values. The answer register is in an evenly weighted superposition of $|0\rangle$ and $|1\rangle$ states. Next, the function $f$ is evaluated using the transformation $U_f : |x,y\rangle \to |x, y \oplus f(x)\rangle$, giving

$$|\psi_2\rangle = \sum_x \frac{(-1)^{f(x)}|x\rangle}{\sqrt{2^n}} \left(\frac{|0\rangle - |1\rangle}{\sqrt{2}}\right). \qquad (3)$$

Using a Hadamard transform on the query register in $|\psi_2\rangle$ we have

$$|\psi_3\rangle = \sum_z \sum_x \frac{(-1)^{x \cdot z + f(x)}|z\rangle}{2^n} \left(\frac{|0\rangle - |1\rangle}{\sqrt{2}}\right). \qquad (4)$$

Assuming that $f$ is constant the amplitude for $|0\rangle^{\otimes n}$ is +1 or -1, depending on the constant value $f(x)$ takes. Because $|\psi_3\rangle$ is of unit length it follows that all other amplitudes must be zero, and an observation will yield 0s for all qubits in the query register.

*2.4 APL2 simulation of the Deutsch-Jozsa algorithm*

All the APL2 code is listed in the Appendix. The Hadamard gate is defined as a global array H by the DEFINEHGATE function. The array H has the following items

```
0.7071067812   0.7071067812
0.7071067812  ⁻0.7071067812
```

This is a computer implementation of $\frac{1}{\sqrt{2}}\begin{bmatrix} 1 & 1 \\ 1 & -1 \end{bmatrix}$. Up to $\frac{1}{\sqrt{2}}$ constant, the APPLY_H_GATE function does the Hadamard gate action on the states $|0\rangle$ or $|1\rangle$ resulting in

```
            APPLY_H_GATE  1
    0  ¯1
            APPLY_H_GATE  0
    0  +1
```

where `0 ¯1` corresponds to $\frac{1}{\sqrt{2}}(|0> - |1>)$ when H acts on $|1>$. Similarly, `0 +1` represents $\frac{1}{\sqrt{2}}(|0> + |1>)$ when H acts on $|0>$ which is expressed by `APPLY_H_GATE 0`. Finally, the HNGATE function simulates, up to a normalization constant, the Hadamard transform on the query register. For example, in the $|\psi_1>$ state and $n = 3$, $\sum_{x \in \{0,1\}^n} \frac{|x>}{\sqrt{2^n}}$ is represented by

```
    000  +001  +010  +011  +100  +101  +110  +111.
```

This is the output for `HNGATE 0 0 0`.

The Deutsch-Jozsa black box $U_f$ which performs transformation $|x>|y> \rightarrow |x>|y \oplus f(x)>$ is simulated in the DJBOX operator. The DJBOX operator simulates essential elements of the quantum circuit for the Deutsch-Jozsa algorithm. In the APL2 sense, an operator is a function that takes another function and variables as an input and perform some calculations. The DJBOX evaluates $f$. $y \oplus f(x)$ is simulated by `Y YPLUSF(F X[I;])` and we assume $f(x) = 1$ for all $x$. $\sum_z \sum_x \frac{(-1)^{x \cdot z + f(x)}}{2^n} |z>$ is represented by an array with $n$ rows and columns. For example, assuming $n = 3$, this array (`P3CONV`) is

```
    0  ¯1  ¯2  ¯3  ¯4  ¯5  ¯6  ¯7
    0   1  ¯2   3  ¯4   5  ¯6   7
    0  ¯1   2   3  ¯4  ¯5   6   7
    0   1   2  ¯3  ¯4   5   6  ¯7
    0  ¯1  ¯2  ¯3   4   5   6   7
    0   1  ¯2   3   4  ¯5   6  ¯7
    0  ¯1   2   3   4   5  ¯6  ¯7
    0   1   2  ¯3   4  ¯5  ¯6   7
```

where the query register states are represented in the decimal representation for a shorter notation. The act of observation is simulated by `+/P3CONV` which is summation of all items in each column of `P3CONV` resulting in `0 0 0 0 0 0 0 0`. This is equivalent to an observation which yields 0s for all qubits in the query register.

*2.5 APL2 simulation of a Vector Quantum Computer that implements the Deutsch-Jozsa algorithm*
      This section develops a simulation of a vector quantum computer that implements the Deutsch-Jozsa algorithm from a machine language view. In this view a computer is a device that has a way of remembering numbers and combining numbers to form new numbers. We think of the computer's memory as an ordered list of individual memory cells, each of which can hold one number. This ordered

list is called main memory. Most of the computers have some specialized high-speed storage areas called registers where most computations are performed. A vector architecture has a set of vector registers, instructions to move values from main memory into vector registers, perform computations on vector registers, and move results back into main memory. In addition, some control registers record the number of items in the vector registers and other related information.

The essentials of the model that represents a modest machine are defined in the DEFINEMACHINE function (see Appendix). A 100-item vector is used to represent main memory. The Deutsch-Jozsa query register size $QR$ is the function parameter. This parameter determines the number of the Deutsch-Jozsa query register states $QRSIZE$ by

$$QRSIZE = 2^{QR}.$$

The four control registers are needed. The first defines the length of each of 5 vector registers and is called the section size $SS$. This number is identical to $QRSIZE$. The second control register $VCT$ records the number of items actually residing in the vector registers. This number can range from zero up to the section size. The third control register $PSS$ determines the number of parts of an array to which a specified function is applied independently. This number can range from zero up to $PSS$. We set $PSS$ equal to $QR$. In addition to 5 vector registers, there are $QRSIZE$ additional vector registers denoted as the Deutsch-Jozsa vector registers. These registers store parts of an array, one part per register, to which a specified function is applied independently. The last control register is a partition vector count register $PVCT$ which records the number of items residing in the Deutsch-Jozsa vector registers.

The memory and registers are initialized to dots (.) so it is easy to distinguish items that have been used from items that have not been used. In a real machine, the initial values would be probably zero. Executing `DEFINEMACHINE 3`, defines the machine with 3 qubits in the Deutsch-Jozsa query register. The SHOMACHINE function (see Appendix) depicts the machine contents:

```
        SHOWMACHINE
MAIN MEMORY                          VCT PVCT

   1 :.........................       0    0
  26 :.........................
  51 :.........................
  76 :.........................

VECTOR REGISTERS

V1:........
V2:........
V3:........
V4:........
V5:........

DJ QUERY REGISTER    DJ VECTOR REGISTERS

Q1:...               V1:................
                     V2:................
                     V3:................
                     V4:................
                     V5:................
                     V6:................
                     V7:................
                     V8:................
```

*2.6 Partitioning in the Vector Quantum Computer that implements the Deutsch-Jozsa algorithm*

The GET_PARTITION function (Appendix) determines parts of an array based on the specified function F3. This function always returns 1 and is evaluated by the Deutsch-Jozsa algorithm which is simulated in the DJBOX operator. For example, for the machine with 3 qubits in the Deutsch-Jozsa query register and the array *A* from the section 2.2, the GET_PARTITION function returns respective partition vector stored in main memory at address 76:

```
       DEFINEMACHINE 3
       GET_PARTITION A[;1]
QUERY REQISTER IN 0 STATE
QUERY REQISTER IN 0 STATE
 MAIN MEMORY                                              VCT PVCT

   1 : . 0 36 61 3 86 . . .  . . . . . . . . . . . . .     0    3
  26 : A A A   A   B B   B C C C  . . . . . . . . . . . . .
  51 : C A A   A   A B   B B C C  . . . . . . . . . . . . .
  76 : 1 0 0   0 1   0 0 1 0 0    . . . . . . . . . . . . .

 VECTOR REGISTERS

 V1: C C A A B B B C
 V2: C C A A A B B B
 V3: 0 0 0 0 1 0 0 1
 V4: . . . . . . . .
 V5: . . . . . . . .

 DJ QUERY REGISTER    DJ VECTOR REGISTERS

  Q1: 0 0 0            V1:. . . . . . . . . . . . . . . .
                       V2:. . . . . . . . . . . . . . . .
                       V3:. . . . . . . . . . . . . . . .
                       V4:. . . . . . . . . . . . . . . .
                       V5:. . . . . . . . . . . . . . . .
                       V6:. . . . . . . . . . . . . . . .
                       V7:. . . . . . . . . . . . . . . .
                       V8:. . . . . . . . . . . . . . . .
```

We assumed those bins into which to accumulate items are already defined outside the F3 function. It is possible to imagine a quite general function F3 which would determine parts (bins) based on constraints of the problem we are trying to solve. The section of the code that determines 0 or 1 to be stored in a vector register cell would be a problem specific. A respective partition vector would be again stored in main memory at address 76.

The CALC_WITH_PARTITIONS function (Appendix) determines parts of an array and then applies a specified function to each independent part. For example, for the machine with 3 qubits in the Deutsch-Jozsa query register and the array *A* from the section 2.2, F2 is the function applied to each independent part. This function is evaluated by the Deutsch-Jozsa algorithm which is simulated in the DJBOX operator. The final result is stored in main memory at address 51:

```
    DEFINEMACHINE 3
        CALC_WITH_PARTITIONS SARRAY
QUERY REQISTER IN 0 STATE
QUERY REQISTER IN 0 STATE
QUERY REQISTER IN 0 STATE
 MAIN MEMORY                                                        VCT PVCT

   1 : .     10   26   76   0   86   54   .   .   .  ...............  0    0
  26 : 100  100   20  400  30  200  300  100 100 100  ...............
  51 : 620  530  300   A    A    B    B    B   C   C  ...............
  76 :   1    0    0   0    1    0    0    1   0   0  ...............

 VECTOR REGISTERS

 V1: C    C    A    A  B  B  B  C
 V2: C    C    A    A  A  B  B  B
 V3: 620  530  300  0  1  0  0  1
 V4: .    .    .    . . . . .
 V5: .    .    .    . . . . .

 DJ QUERY REGISTER   DJ VECTOR REGISTERS

  Q1: 0 0 0          V1: 4 100 100   20 400 ...........
                     V2: 3  30 200  300  .  ...........
                     V3: 3 100 100  100  .  ...........
                     V4: . .   .    .    .  ...........
                     V5: . .   .    .    .  ...........
                     V6: . .   .    .    .  ...........
                     V7: . .   .    .    .  ...........
                     V8: . .   .    .    .  ...........
```

In this example the F2 function sums elements of each partition. Obviously the section of the code that applies a specified function to each partition is a problem specific and needs to be rewritten for every problem we are trying to solve. Note that in the first part of the CALC_WITH_PARTITIONS function we determine a respective partition vector so we could use a function call to the GET_PARTITION function to do this task instead of explicit coding.

**(3) Discussion**

We used the Deutsch-Jozsa algorithm to solve a class of problems in which we need to determine parts of an array and then apply a specified function to each independent part. Since present quantum computers are not robust enough for code writing and execution, we developed an APL2 model of a vector quantum computer that implements the Deutsch-Jozsa algorithm from a machine language view. The model is based on the example of the vector architecture from the reference [3]. It requires two additional vector register sets: the Deutsch-Jozsa query register and the Deutsch-Jozsa vector registers.

The Deutsch-Jozsa query register corresponds to the query register of the Deutsch-Jozsa algorithm. The Deutsch-Jozsa vector registers store parts of an array, one part per register, to which a specified function is applied independently. The array parts and respective partition vector are determined in the remaining vector registers. The totality of the vector registers forms a framework for solving the above class of problems.

Two additional control registers are needed. One control register determines the number of parts of an array to which a specified function is applied independently. The other is a partition vector count register which records the number of items residing in the Deutsch-Jozsa vector registers.

We know that the function $f$ evaluated by the Deutsch-Jozsa algorithm is either constant for all inputs or else $f$ is balanced, that is, equal to 1 for exactly half of all possible inputs. The Deutsch-Jozsa algorithm produces an answer that is always correct with just 1 evaluation of $f$. In our model we assume $f(x) = 1$ for all $x$. $f(x)$, before returns 1, does one of two possible computations while it is evaluated by the Deutsch-Jozsa algorithm. $f(x)$ determines parts of an array or it applies a specified function to each independent part. For example, F3 function when evaluated by the DJBOX operator compares two vector register cell contents and store the result in the third vector register cell. The address of all three cells is equal to the decimal representation of the query register state $x$. For example, if $x = 100$, the address is equal to 4. Similarly, F2 function sums elements of the vector stored in a Deutsch Jozsa vector register and stores the sum in the corresponding vector register. The addresses of the respective cells are constrained by the decimal representation of the query register state $x$. Therefore, both the length of the vector registers and the number of the Deutsch-Jozsa vector registers are equal to the number of the query register states ($QRSIZE$).

The DJBOX operator is the core of the method, since the Deutsch-Jozsa algorithm needs to evaluate an arbitrary function $f$ which returns always 1. The operator is the simplest design choice to "an arbitrary function $f$" condition since the operator is a function that takes another function and variables as an input and perform some calculations. The operator concept is a part of the APL2 syntax, thus the APL2 seems a natural language for the model of a vector quantum computer that implements the Deutsch-Jozsa algorithm from a machine language view. An additional bonus of implementing the APL2 on a quantum computer is that the APL2 deals with whole collections of data all at once. This is similar to quantum parallelism which enables all possible values of a function to be evaluated simultaneously, even we apparently only evaluated the function once.

It was shown [4] that computer languages can be reduced to an algebraic representation. By algebraic we mean that the computer language can be represented with operator expressions using operators that have an algebra similar to that of the raising and lowering operators in quantum mechanics. To establish the algebraic representation we associate a harmonic raising operator $a_i^+$ and a lowering operator $a_i$ with each memory location $i$. These operators satisfy the commutation relations

$$[a_i, a_j^+] = \delta_{ij}$$
$$[a_i, a_j] = 0$$
$$[a_i^+, a_j^+] = 0$$

where $\delta_{ij}$ is 1 if $i = j$ and zero otherwise. We also define a pair of raising and lowering operators for the register $r$ and $r^+$ with commutation relations

$$[r_i, r_j^+] = \delta_{ij}$$
$$[r_i, r_j] = 0$$
$$[r_i^+, r_j^+] = 0.$$

The ground state of the quantum computer is the state with the values at all memory locations set to zero. It is represented by the vector $|0,0,0,...\rangle$. A general state of the quantum computer will be represented by a vector of the form

$$|n,m,p,...\rangle = N(r^+)^n (a_0^+)^m (a_1^+)^p ... |0,0,0,...\rangle$$

where $N$ is a normalization constant and with the first number being the value in the register, the second number the value at memory location 0, the third number the value at memory location 1, and so on.

Let us imagine that the APL2 is, indeed, implemented on a quantum computer with the vector architecture. All the APL2 statements are represented with the above raising and lowering operators. Then the code for both functions GET_PARTITION and CALC_WITH_PARTITIONS could be executed, without any modification, on such a computer. On the other hand, the DJBOX operator just simulates the Deutsch Jozsa black box, since the code is looping by the query register states:

```
L0:->(QRSIZE<I←I+1)↑LX ⍝ LOOP BY THE QUERY REGISTER STATES
 LOAD_QUERY_REGISTER X[I;] ⍝ LOAD THE QUERY REGISTER WITH STATE X[I;]
⍝ NOTE: X IS AN ARRAY OF ALL SUPERPOSITION VALUES IN THE QUERY REGISTER
 T←HNGATE X[I;] ⍝ APPLY HADAMARD GATE ON THE STATE X[I;]
 P←~∊(T='+')∨(T='-') ⍝ IDENTIFY WHERE IS NOT + OR - (FIND STATES)
 TT←P⊂∊T ⍝ ENCLOSE STATES
 CONV←2⊥⍣¨⊃TT ⍝ EXPRESS EACH STATE IN DECIMAL REPRESENTATION (STATES ARE IN BINARY REPRESENTATION)
 Z←(~P)/∊T ⍝ COLLECT ALL +,-
 K←0 ⍝ INITIALIZE INNER LOOP VARIABLE
 TTT←'' ⍝ INITIALIZE VARIABLE WHICH HOLDS RESULT OF LOOP ITERATION
L1:->((QRSIZE-1)<K←K+1)↑L2 ⍝ LOOP QRSIZE-1 TIMES, BEGIN LOOP WITH THE SECOND STATE
 TTT←TTT,(1+Z[K]='+')⊃(¯1×CONV[1+K])(CONV[1+K])⍝ IF + SIGN ADD STATE,
⍝ IF - SIGN, ADD STATE MULTIPLIED BY ¯1
 ->L1 ⍝ GO TO NEXT LOOP ITERATION
L2: ⍝ EXIT LOOP
 P3CONV←P3CONV,[1](Y YPLUSF(F X[I;]))×(0,TTT) ⍝ ADD 0 STATE TO TTT,
⍝ MULTIPLY STATES WITH THE OUTPUT OF Y ADD MODULO 2 F(X) AND
⍝ ADD THEM TO P3CONV
 ->L0 ⍝ GO TO NEXT STATE IN THE QUERY REGISTER
LX:⍝ EXIT OUTER LOOP
```

Respective quantum computer code would take advantage of the quantum superposition and entanglement principles instead of looping and we could pack this code into a new machine code statement. The statement would replace above looping by the query register states. Such a modified DJBOX operator could be executed on the quantum computer with the vector architecture.

### (4) Conclusions

In this paper we developed, using the Deutsch-Jozsa algorithm, a method to solve a class of problems in which we need to determine parts of an array and then apply a specified function to each independent part. To evaluate the method, since present quantum computers are not robust enough for code writing and execution, we built an APL2 model of a vector quantum computer that implements the Deutsch-Jozsa algorithm from a machine language view. The core of the method is an operator which allows evaluation of an arbitrary function $f$ by the Deutsch-Jozsa algorithm The operators' concept is already a part of the APL2 syntax. Although the APL2 dealing with whole collections of data all at once is similar to quantum parallelism, it is founded on an implicit looping through the data. Quantum parallelism does not need any consecutive processing. Thus the APL2 implemented on a vector quantum computer would have statements represented with operator expressions using operators that have an algebra similar to that of the raising and lowering operators in quantum mechanics. Taking advantage of the quantum superposition and entanglement principles, such an APL2 ("quantum APL2") would be significantly more powerful and efficient than its classical version. We outlined how to adapt the DJBOX operator code so two key functions for the solution of the above class of the problems (GET_PARTITION, CALC_WITH_PARTITIONS) can be executed on the vector quantum computer. Once when the vector quantum computer with "quantum APL2" is robust enough for code writing and execution, the above method could be implemented and further explored.

A novel feature of our model of the vector quantum computer is the Deutsch-Jozsa vector registers. These registers are exclusively used for the Deutsch-Jozsa algorithm execution. Thus, the Deutsch-Jozsa algorithm is an integral part of the vector quantum computer architecture. Although we assumed a constant function $f$ evaluated by the Deutsch-Jozsa algorithm, our method can be easily generalized for a balanced function too.

In this paper we were not concerned whether we have designed the most efficient vector quantum computer, nor how we could best implement it. We wanted to show that it is possible, in principle, to use the Deutsch-Jozsa algorithm as an integral part of the vector quantum architecture to solve the class of problems in which we need to determine parts of an array and then apply a specified function to each independent part. Since the Deutsch-Jozsa algorithm takes advantage of the quantum superposition and entanglement principles, we believe our method would be more powerful and efficient than its classical version.

**Acknowledgments**

The results presented on these pages are the outcome of independent research not supported by any institution or government grant.

## Appendix

DEFINEMACHINE function defines global variables of the vector machine.

```
DEFINEMACHINE A;⎕IO
⍝ DESCRIPTION:DEFINE VECTOR MACHINE GLOBALS
 ⎕IO←1 ⍝ ORIGIN
⍝ INPUT:A DEFINES Deutsch-JOZSA QUERY REGISTER SIZE

⍝ DEFINE GLOBAL MEMORY AREAS
 MM←100⍴'.' ⍝ MAIN MEMORY
 QR←A ⍝ SET Deutsch-JOZSA QUERY REGISTER SIZE
 QRSIZE←2*QR    ⍝ NUMBER OF Deutsch-JOZSA QUERY REGISTER STATES
 SS←QRSIZE      ⍝ SECTION SIZE
 PSS←QR         ⍝ PARTITION SECTION SIZE
 VR←(5 SS)⍴'.'  ⍝ 5 VECTOR REGISTERS
 VCT←0          ⍝ VECTOR COUNT REGISTER
 PVCT←0         ⍝ PARTITION VECTOR COUNT REGISTER
 DJ←(1 QR)⍴'.'  ⍝ Deutsch-JOZSA QUERY REGISTER
⍝ Deutsch-JOZSA VECTOR REGISTERS
 DJVR←(QRSIZE(2×QRSIZE))⍴'.'
```

SHOMACHINE function displays contents of the vector machine

```
SHOWMACHINE;⎕IO;R;N
⍝ DESCRIPTION: DEPICTS THE MACHINE'S CONTENTS
 ⎕IO←1 ⍝ ORIGIN
⍝ INPUT: NONE

⍝ DEPICT MAIN MEMORY, VECTOR
⍝ AND PARTITION VECTOR COUNT REGISTERS
 N←'MAIN MEMORY' 'VCT' 'PVCT'
 R←↑(⍴MM)÷25 ⍝ DEFINE 4 ROWS FOR MAIN MEMORY
 N,[0.5](⍕ 1⍴1+25×⁻1+⍳R),':',R 25⍴MM)VCT PVCT
 ''
⍝ DEPICT VECTOR REGISTERS
 N←'VECTOR REGISTERS' ' '
 N,[0.5](('V',⍕5 1⍴⍳5),':',VR)(' ')
 ''
⍝ DEPICT Deutsch JOZSA QUERY AND VECTOR
⍝ REGISTERS
 N←'DJ QUERY REGISTER' 'DJ VECTOR REGISTERS'
 N,[0.5](('Q',⍕1 1⍴1),':',DJ)(('V',⍕QRSIZE 1⍴⍳QRSIZE),':',DJVR)
```

SETMEMORY function loads data V into the main memory at address A..

```
R←A SETMEMORY V;⎕IO
⍝ DESCRIPTION: PUT V INTO MAIN MEMORY AT ADDRESS A
 ⎕IO←1 ⍝ ORIGIN
⍝ INPUT: A MAIN MEMORY ADDRESS
⍝        V DATA TO PUT INTO MAIN
⍝          MEMORY

 V←,V ⍝ IGNORE SHAPE OF DATA
 ((⍴V)↑(A-1)↓MM)←V ⍝ MOVE DATA TO MEMORY
```

LOADV function loads vector register V from address in A with stride S.

```
LOADV R;⎕IO;V;A;S
⍝ DESCRIPTION: LOAD VECTOR REGISTER V FROM ADDRESS
⍝ IN A WITH STRIDE S
 ⎕IO←1  ⍝ ORIGIN
⍝ INPUT: R( 3 ITEMS VECTOR(V, A, S))
⍝        V IS VECTOR REGISTER
⍝        A IS MEMORY ADDRESS
⍝        S IS STRIDE

 (V A S)←R ⍝ DESCOMPOSE R INTO V, A, AND S
⍝ LOAD VECTOR REGISTER V FROM ADDRESS IN A WITH STRIDE S
 VR[V;⍳VCT]←MM[MM[A]+S×¯1+⍳VCT]
⍝ UPDATE MAIN MEMORY WITH THE LENGTH OF LOADED DATA
 MM[A]←MM[A]+VCT
```

STOREV function stores vector register V in address in A with stride S into main memory.

```
STOREV R;⎕IO;V;A;S;P
⍝ DESCRIPTION: STORE VECTOR REGISTER IN
⍝ ADDRESS IN A WITH STRIDE S INTO MAIN MEMORY
 ⎕IO←1  ⍝ ORIGIN
⍝ INPUT: R( 3 ITEMS VECTOR(V, A, S))
⍝        V IS VECTOR REGISTER
⍝        A IS MEMORY ADDRESS
⍝        S IS STRIDE
 (V A S)←R
⍝ STORE VECTOR REGISTER V IN ADDRESS IN A
⍝ WITH STRIDE S INTO MAIN MEMORY
 MM[MM[A]+S×¯1+⍳VCT]←VR[V;⍳VCT]
⍝ UPDATE MAIN MEMORY WITH THE LENGTH OF STORED DATA
 MM[A]←MM[A]+VCT
```

LOADVCT function loads vector count from memory address A.

```
LOADVCT A;⎕IO
⍝ DESCRIPTION: LOAD VECTOR COUNT FROM
⍝ MEMORY ADDRESS A
 ⎕IO←1  ⍝ ORIGIN
⍝ INPUT: MEMORY ADDRESS A

⍝ LOAD VECTOR COUNT FROM MEMORY ADDRESS A
 VCT←SS⌊MM[A] ⍝ VECTOR COUNT IS MINIMUM
⍝ BETWEEN SECTION SIZE (SS) AND VALUE IN MAIN MEMORY AT ADDRESS A
 MM[A]←MM[A]-VCT ⍝ REDUCE STORAGE COUNT
```

LOADPVCT function loads partition vector count from memory address A.

```
LOADPVCT A;⎕IO
⍝ DESCRIPTION: LOAD PARTITION VECTOR COUNT FROM
⍝ MEMORY ADDRESS A
 ⎕IO←1  ⍝ ORIGIN
⍝ INPUT: MEMORY ADDRESS A

⍝ LOAD PARTITION VECTOR COUNT FROM MEMORY ADDRESS A
 PVCT←MM[A]
 MM[A]←0 ⍝ REDUCE STORAGE COUNT TO 0
```

PSTOREV function stores vector register V that contains results of applying a function to partitions into main memory at address A with stride S.

```
PSTOREV R;⎕IO;V;A;S
⍝ DESCRIPTION: STORE VECTOR REGISTER
⍝ THAT CONTAINS RESULTS OF APPLYING A FUNCTION
⍝ TO PARTITIONS INTO MAIN MEMORY AT ADDRESS A
⍝ WITH STRIDE S
 ⎕IO←1 ⍝ ORIGIN
⍝ INPUT: R( 3 ITEMS VECTOR(V, A, S))
⍝        V IS VECTOR REGISTER
⍝        A IS MAIN MEMORY ADDRESS
⍝        S IS STRIDE

 (V A S)←R
⍝ STORE VECTOR REGISTER V INTO MAIN MEMORY
⍝ AT ADDRESS A WITH STRIDE S
 MM[MM[A]+S×⁻1+⍳PVCT]←VR[V;⍳PVCT]
⍝ UPDATE MAIN MEMORY WITH THE LENGTH OF STORED DATA
 MM[A]←MM[A]+PVCT
```

DEFINEHGATE function defines the Haddamard gate as global variable H.

```
DEFINEHGATE
⍝ DESCRIPTION: DEFINE HADDAMARD GATE(H)
 ⎕IO←1 ⍝ ORIGIN
⍝ OUTPUT: GLOBAL ARRAY H

⍝ DEFINE 2 BY 2 ARRAY WITH ELEMENTS
⍝ 1   1
⍝ 1  ⁻1
 H←2 2⍴1 1 1 ⁻1
⍝ MULTIPLY WITH 1/SQRT(2)
 H←(1÷2*0.5)×H
```

APPLY_H_GATE function simulates, up to $\frac{1}{\sqrt{2}}$ constant, the Hadamard gate action on states $|0>$ or $|1>$.

```
 R←APPLY_H_GATE A;⎕IO;Q1;Q2;S;O;O1;O2;TO1;TO2
⍝ DESCRIPTION: SIMULATES ACTION OF THE
⍝ HADAMARD GATE(H) ON STATES |0> OR |1>
⍝ NEGLECTING SQRT(1/2) NORMALIZATION CONSTANT
⍝ WHEN H ACTS ON |0>, THE FUNCTION RETURNS (|0>+|1>) STATE
⍝ LABELED AS '0' '+1'
⍝ WHEN H ACTS ON |1>, THE FUNCTION RETURNS (|0>-|1>) STATE
⍝ LABELED AS '0' '-1'
 ⎕IO←0 ⍝ ORIGIN
⍝ INPUT: 1 (SIMULATES STATE |1>)
⍝        0 (SIMULATES STATE |0>)
 Q1←2 1⍴1 0 ⍝ |0> STATE
 Q2←2 1⍴0 1 ⍝ |1> STATE
 S←(A=1)⊃Q1 Q2 ⍝ IF A = 1 ASSIGN |1> STATE TO S
⍝ OTHERWISE IF A = 0 ASSIGN |0> STATE TO S
 O←H+.×S    ⍝ H ACTS ON S
 O1←(1÷2*0.5)×(Q1+Q2) ⍝ OUTPUT WHEN H ACTS ON |0>
 TO1←⊂'0' '+1' ⍝ THIS WILL BE RETURNED
 O2←(1÷2*0.5)×(Q1-Q2) ⍝ OUTPUT WHEN H ACTS ON |1>
 TO2←⊂'0' '-1' ⍝ THIS WILL BE RETURNED
 R←(∧/O=O1)⊃(TO2)(TO1) ⍝ IF O = O1 RETURN TO1 OTHERWISE RETURN TO2
```

HNGATE function simulates, up to a normalization constant, the Hadamard transform on the query register.

```
R←HNGATE V;⎕IO;T;C;BP;BM;S;FS;I
⍝ DESCRIPTION: SIMULATES UP TO NORMALIZATION CONSTANT
⍝ HADAMARD TRANSFORM ON THE QUERY REGISTER
 ⎕IO←1 ⍝ ORIGIN
⍝ INPUT: QUERY REGISTER STATE(V)
⍝ FOR EXAMPLE IF THE QUERY REGISTER DESCRIBES
⍝ THE STATE OF 3 QUBITS ALL PREPARED IN THE |0> STATE,
⍝ V IS 0 0 0

 DEFINEHGATE ⍝ CREATE HADAMARD ARRAY H
 T←APPLY_H_GATE¨V ⍝ APPLY H ON EACH COMPONENT OF V AND
⍝ STORE RESULT IN T
 →(0=⍴⍴T)↓L0 ⍝ CHECK IF T HAS ONLY ONE COMPONENT WHICH
⍝ INDICATES THAT V HAS ONLY ONE STATE(0 OR 1); IF NOT
⍝ JUMP TO LABEL L0
 R←T ⍝ STORE T IN FINAL RESULT
 →LX ⍝ RETURN R
L0: ⍝ V HAS MORE THAN 1 STATE WHICH MEANS T HAS MORE THAN
⍝ ONE COMPONENT
 FS←T[1]CALC_SIGN T[2]⍝ CALCULATE SIGNS FOR THE TERMS IN TENSOR
⍝ PRODUCT OF STATES FOR THE FIRST 2 COMPONENTS OF T
⍝ + SIGN IS REPRESENTED BY 1
⍝ - SIGN IS REPRESENTED BY ¯1
 R←T[1]COMPOSE_STATE T[2]⍝ NEGLECTING NORMALIZATION CONSTANT
⍝ DO TENSOR PRODUCT OF STATES FOR THE FIRST 2 COMPONENTS OF T
⍝ NOTE: COMPOSE_STATE USES FS VARIABLE
 →(2=⍴V)↑LX ⍝ CHECK IF V HAS ONLY 2 STATES; IF SO
⍝ RETURN R
 I←2 ⍝ INITIALIZE LOOP VARIABLE
⍝ V HAS MORE THAN 2 STATES
⍝ LOOP THROUGH THE REST OF T COMPONENTS
L1:→((⍴T)<I←I+1)↑LX ⍝ START WITH I=3; RETURN FINAL R WHEN
⍝ THE REST OF T COMPONENTS ARE PROCESSED
 FS←R CALC_SIGN T[I]⍝ CALCULATE SIGNS FOR THE TERMS IN TENSOR
⍝ PRODUCT BETWEEN R AND I-TH COMPONENT OF T
 R←R COMPOSE_STATE T[I] ⍝ NEGLECTING NORMALIZATION CONSTANT
⍝ DO TENSOR PRODUCT BETWEEN R AND I-TH COMPONENT OF T
 →L1 ⍝ GO BACK TO THE START OF THE LOOP
LX:⍝ EXIT
```

For example, below is displayed the output of the HNGATE function when the query register describes the state of 3 qubits all prepared in the |0> state.

```
      HNGATE 0 0 0
 000 +001 +010 +011 +100 +101 +110 +111
```

CALC_SIGN function calculates +,- signs for terms in the tensor product of 2 quantum states.

```
R←A CALC_SIGN B;⎕IO;C;BP;BM;S;L
⍝ DESCRIPTION: CALCULATE SIGNS FOR THE TERMS IN TENSOR
⍝ PRODUCT OF STATES A AND B
⍝ THIS FUNCTION IS CALLED FROM HNGATE FUNCTION
 ⎕IO←1  ⍝ ORIGIN
⍝ INPUT: STATES A AND B

 L←(1+0=⍴⍴↑⊃A)⊃(↑⊃A)(A)  ⍝ IF A
⍝ IS ONE COMPONENT OF THE APPLY_H_GATE FUNCTION OUTPUT,
⍝ ASSIGN ↑⊃A TO L SINCE DEPTH OF A IS 4, OTHERWISE IF A IS ONE COMPONENT OF R
⍝ WHICH IS BUILD IN HNGATE FUNCTION, ASSIGN A TO L SINCE DEPTH OF A IS 2
⍝ DEPTH OF L IN BOTH CASES IS 2
 C←,L∘.,(↑⊃B)  ⍝ CATENATE EVERY ELEMENT OF L TO EVERY ELEMENT OF
⍝ ↑⊃B, SINCE DEPTH OF B IS 4; DEPTH OF ↑⊃B IS 2;
⍝ L∘.,(↑⊃B) IS AN ARRAY SO WE ROLL IT INTO A VECTOR C
 BP←C='+'  ⍝ IDENTIFY WHERE C HAS + SIGN (VALUE OF BP IS 1);
⍝ WHERE C DOES NOT HAVE + SIGN, VALUE OF BP IS 0
 BM←C='-'  ⍝ IDENTIFY WHERE C HAS - SIGN (VALUE OF BM IS 1);
⍝ WHERE C DOES NOT HAVE - SIGN, VALUE OF BM IS 0
 S←BP+¯1×BM  ⍝ MULTIPLY BM WITH ¯1 SINCE BM SHOWS WHERE C HAS - SIGN AND
⍝ ADD TO BP; S STILL HAS ZEROS WHERE C IS NOT + OR -
 S←S+S=0  ⍝ IDENTIFY WHERE S HAS ZEROS (VALUE OF S=0 IS 1 IN THAT CASE;
⍝ OTHERWISE IS 0) AND ADD S=0 TO S;
⍝ THE FINAL S HAS ONLY 1 OR ¯1
⍝ DEPTH OF S IS 2
 R←×/¨S  ⍝ MULTIPLY ELEMENTS IN EACH PART OF S AND ASSIGN
⍝ PRODUCT FOR EACH PART TO R; WHERE THE FINAL SIGN IS +, R HAS VALUE OF 1
⍝ WHERE THE FINAL SIGN IS -, R HAS VALUE OF ¯1
```

COMPOSE_STATE function calculates the tensor product of 2 quantum states.

```
R←A COMPOSE_STATE B;⎕IO;C;PM;V;M;PV;FV;I;L
⍝ DESCRIPTION:NEGLECTING NORMALIZATION CONSTANT
⍝ CALCULATE TENSOR PRODUCT OF 2 STATES
⍝ THIS FUNCTION IS CALLED FROM THE HNGATE FUNCTION
 ⎕IO←1  ⍝ ORIGIN
⍝ INPUT: STATES A AND B
⍝        GLOBAL VARIABLE FS DEFINED IN HNGATE FUNCTION

 L←(1+0=⍴⍴↑⊃A)⊃(↑⊃A)(A)⍝ IF A
⍝ IS ONE COMPONENT OF THE APPLY_H_GATE FUNCTION OUTPUT,
⍝ ASSIGN ↑⊃A TO L SINCE DEPTH OF A IS 4, OTHERWISE IF A IS ONE COMPONENT OF R
⍝ WHICH IS BUILD IN HNGATE FUNCTION, ASSIGN A TO L SINCE DEPTH OF A IS 2
⍝ DEPTH OF L IN BOTH CASES IS 2
 C←,L∘.,(↑⊃B)⍝ CATENATE EVERY ELEMENT OF L TO EVERY ELEMENT OF
⍝ ↑⊃B, SINCE DEPTH OF B IS 4; DEPTH OF ↑⊃B IS 2;
⍝ L∘.,(↑⊃B) IS AN ARRAY SO WE ROLL IT INTO A VECTOR C
 PM←(C='+')∨(C='-')⍝ IDENTIFY WHERE VALUE OF C IS +
⍝ OR - (VALUE OF PM IS 1); OTHERWISE VALUE OF PM IS 0
 V←(∊~PM)/∊C  ⍝ SELECT ALL OTHER VALUES OF C (NOT EQUAL TO + OR -)
 M←(⍴∊C[1])⍴0  ⍝ CREATE VECTOR M THAT HAS ALL COMPONENTS EQUAL TO 0
⍝ M AND C[1] HAVE THE SAME NUMBER OF ELEMENTS
 M[1]←1  ⍝ SET THE FIRST ELEMENT EQUAL TO 1
 PV←((⍴C)×⍴∊C[1])⍴M  ⍝ CREATE A PARTITION VECTOR;
⍝ PV IS EQUAL TO 1 AT THE FIRST ELEMENT OF EVERY COMPONENT OF C
⍝ OTHERWISE IS 0
 FV←(+\PV)⊂V  ⍝ PARTITION V USING PV
 R←FV[1]  ⍝ ASSIGN THE FIRST COMPONENT OF FV TO R
 I←1  ⍝ INITIALIZE LOOP COUNTER
L0:->((⍴FV)<I←I+1)↑LX   ⍝ START WITH I=2 AND LOOP THROUGH FV COMPONENTS
 R←R,⊂∊((1+FS[I]=¯1)⊃'+' '-'),FV[I]⍝ USING GLOBAL FS DETERMINE
⍝ WHICH SIGN TO USE (IF FS[I] = ¯1 TAKE -, OTHERWISE TAKE +) AND
⍝ CATENATE THE SIGN AND FV[I] TO R
 ->L0  ⍝ GO BACK TO THE START OF THE LOOP
LX:⍝ EXIT
```

LOAD_QUERY_REGISTER function loads the query register with a vector.

```
LOAD_QUERY_REGISTER A;⎕IO
⍝ DESCRIPTION: LOADS QUERY REGISTER WITH VECTOR A
 ⎕IO←1  ⍝ ORIGIN
⍝ INPUT: VECTOR A

⍝ LOAD QUERY REGISTER WITH VECTOR A
 DJ[;⍳(⍴A)]←A
```

LOAD_PARTITIONS_INTO_DJVR function loads the Deutsch-Jozsa vector registers with partitions.

```
 P LOAD_PARTITIONS_INTO_DJVR L;⎕IO;I
⍝ DESCRIPTION: LOADS Deutsch-JOZSA VECTOR REGISTERS WITH PARTITIONS
 ⎕IO←1  ⍝ ORIGIN
⍝ INPUT: NESTED VECTOR P OF PARTITIONS
⍝        VECTOR L THAT HAS LENGTH OF EACH PARTITION

 I←0  ⍝ SET LOOP COUNTER TO 0
L0:->((⍴P)<I←I+1)↑L1  ⍝ LOOP BY PARTITIONS; WHEN ALL PARTITIONS ARE
⍝ PROCESSED EXIT AT L1
 DJVR[I;1+⍳∊L[I]]←∊P[I]  ⍝ STORE I-TH PARTITION TO I-TH DJ VECTOR REGISTER
⍝ STARTING AT POSITION 2
 DJVR[I;1]←∊L[I]  ⍝ STORE LENGTH OF I-TH PARTITION TO I-TH DJ VECTOR REGISTER
⍝ AT POSITION 1
 ->L0  ⍝ GO TO L0 TO LOAD NEXT PARTITION
L1:  ⍝ EXIT
```

CALC_PARTITIONS function partitions a data vector using given partition vector.

```
 R←CALC_PARTITIONS V;⎕IO;L;A;B;S;P;LN
⍝ DESCRIPTION: PARTITION DATA VECTOR USING GIVEN PARTITION VECTOR
 ⎕IO←1  ⍝ ORIGIN
⍝ INPUT: NESTED VECTOR V WHICH HAS 4 COMPONENTS:
⍝        (MAIN MEMORY ADDRESSES OF 3 ITEMS BELOW)
⍝        1. DATA VECTOR LENGTH
⍝        2. DATA VECTOR
⍝        3. PARTITION VECTOR
⍝        4. STRIDE

 (L A B S)←V  ⍝ PARSE VECTOR V COMPONENTS INTO 4 VARIABLES:
⍝   L IS DATA VECTOR LENGTH MAIN MEMORY ADDRESS
⍝   A IS DATA VECTOR MAIN MEMORY ADDRESS
⍝   B IS PARTITION VECTOR MAIN MEMORY ADDRESS
⍝   S IS STRIDE
 P←(+\MM[MM[B]+S×¯1+⍳MM[L]])⊂MM[MM[A]+S×¯1+⍳MM[L]]  ⍝ PARTITION DATA VECTOR
⍝ AT A WITH LENGTH AT L AND STRIDE S USING PARTITION VECTOR AT B WITH
⍝ LENGTH AT L AND STRIDE S
 LN←⍴¨P   ⍝ STORE LENGTH OF EACH PARTITION IN LN
 R←P LN   ⍝ RETURN PARTITIONS AND RESPECTIVE LENGTHS
```

COUNT_PARTITIONS function calculates how many partitions there are in a partition vector.

```
COUNT_PARTITIONS V;⎕IO;A;S;T
⍝ DEFINITION: CALCULATES HOW MANY PARTITIONS THERE ARE
⍝ IN A PARTITION VECTOR
 ⎕IO←1 ⍝ ORIGIN
⍝ INPUT: NESTED VECTOR V WITH 2 COMPONENTS
⍝        1. MAIN MEMORY ADDRESS WHICH STORES PARTITION VECTOR MAIN MEMORY ADDRESS
⍝        2. STRIDE
⍝ GLOBAL: PVCT (NUMBER OF PARTITIONS)
⍝ THIS FUNCTION IS CALLED FROM CALC_WITH_PARTITIONS FUNCTION

 (A S)←V ⍝ PARSE VECTOR V INTO 2 VARIABLES A AND S:
⍝         A IS MAIN MEMORY ADDRESS WHICH STORES PARTITION VECTOR MAIN MEMORY ADDRESS
⍝         S IS STRIDE
 PVCT←⌈/+\MM[(MM[A]-PVCT)+S×¯1+⍳PVCT] ⍝ COUNT
⍝ PARTITIONS IN THE PARTITION VECTOR AT MAIN MEMORY
⍝ ADDRESS (MM[A]-PVCT) WITH STRIDE S
⍝ NOTE: WE HAVE TO SUBTRACT PVCT FROM MM[A] SINCE
⍝ THE PARTITION VECTOR WAS CALCULATED IN A PREVIOUS
⍝ STEP(FUNCTION CALC_WITH_PARTITIONS) SO THE CONTENT OF THE MAIN MEMORY AT A, MM[A],
WAS
⍝ CHANGED TO MM[A]+PVCT
```

LOADLPVCT function loads a partition vector count from an address in the main memory while looping through partitions and loading them in the Deutsch-Jozsa vector registers.

```
LOADLPVCT A;⎕IO
⍝ DESCRIPTION: LOAD PARTITION VECTOR COUNT FROM MEMORY ADDRESS A
⍝ WHILE LOOPING THROUGH PARTITIONS AND LOADING THEM IN Deutsch-JOZSA VECTOR
⍝ REGISTERS
 ⎕IO←1 ⍝ ORIGIN
⍝ INPUT: MEMORY ADDRESS A

⍝ LOAD PARTITION VECTOR COUNT FROM MEMORY ADDRESS A
 PVCT←PSS⌊MM[A] ⍝ COUNT IS MIN BETWEEN
⍝ PARTITION SECTION SIZE (PSS) AND VALUE IN MAIN MEMORY AT ADDRESS A
 MM[A]←MM[A]-PVCT ⍝ REDUCE STORAGE COUNT
```

PREPARE_T_PARTITIONS function returns a specified number of partitions and respective partition lengths.

```
R←PREPARE_T_PARTITIONS V;⎕IO;LP;P;LN;T;B
⍝ DESCRIPTION: RETURNS THE MOST PSS PARTITIONS
⍝ WITH RESPECTIVE PARTITION LENGTHS
⍝ THIS FUNCTION IS CALLED FROM CALC_WITH_PARTITIONS FUNCTION
⍝ GLOBAL: PSS (PARTITION SECTION SIZE)
 ⎕IO←1 ⍝ ORIGIN
⍝ INPUT: NESTED VECTOR V WITH 3 COMPONENTS:
⍝        1. PARTITIONS LOOP COUNTER (WE ARE LOOPING THROUGH PARTITIONS IN
⍝                                    CALC_WITH_PARTITIONS FUNCTION)
⍝        2. PARTITIONS
⍝        3. RESPECTIVE PARTITION LENGTHS

 (LP P LN)←V ⍝ PARSE OUT COMPONENTS OF V INTO 3 VARIABLES:
⍝          LP  IS PARTITIONS LOOP COUNTER
⍝          P   CONTAINS PARTITIONS
⍝          LN  CONTAINS RESPECTIVE PARTITION LEGNTHS
 T←(LP×PSS)+⍳PSS ⍝ DEFINE PSS INDICES
 B←T≤⍴P  ⍝ ONLY SELECT INDICES LESS THEN OR EQUAL TO THE
⍝ TOTAL NUMBER OF PARTITIONS
 R←P[B/T]LN[B/T] ⍝ RETURN PARTITIONS AND RESPECTIVE PARTITION LENGTHS FOR DEFINED
⍝ INDICES
```

GET_PARTITION function calculates a partition vector.

```
 GET_PARTITION V;⎕IO;L
⍝ DESCRIPTION: CALCULATES PARTITION VECTOR
 ⎕IO←1 ⍝ ORIGIN
⍝ INPUT: GROUP BY (BINS) VECTOR V

⍝ LOAD MAIN MEMORY
 26 SETMEMORY V ⍝ LOAD V IN MAIN MEMORY AT ADDRESS 26
 51 SETMEMORY ¯1⌽V ⍝ LOAD ROTATED V IN MAIN MEMORY AT ADDRESS 51
⍝ ROTATED V HAS THE LAST ELEMENT OF V AT THE FIRST POSITION
 L←⍴V ⍝ FIND LENGTH OF V
 2 SETMEMORY L ⍝ LOAD L IN MAIN MEMORY AT ADDRESS 2
 3 SETMEMORY 26⍝ MAIN MEMORY ADDRESS OF THE VECTOR V
 4 SETMEMORY 51⍝ ¯1⌽V MAIN MEMORY ADDRESS
 5 SETMEMORY L ⍝ PARTITION COUNT ADDRESS IN MAIN MEMORY
 6 SETMEMORY 76⍝ PARTITION VECTOR ADDRESS IN MAIN MEMORY
 LOADPVCT 5 ⍝ LOAD PARTITION VECTOR COUNT
LOOP:LOADVCT 2 ⍝ LOAD VECTOR COUNT
 →(0=VCT)↑LX ⍝ EXIT IF VECTOR COUNT IS ZERO
 LOADV 1 3 1 ⍝ LOAD V1 VECTOR REGISTER WITH V
 LOADV 2 4 1 ⍝ LOAD V2 VECTOR REGISTER WITH ¯1⌽V
 1(F3 DJBOX)0 ⍝ USE DJBOX OPERATOR TO CALCULATE
⍝ PARTITION VECTOR FOR V
 STOREV 3 6 1 ⍝ STORE PARTITION VECTOR IN MAIN MEMORY
 →LOOP ⍝ LOOP BY VECTOR COUNT
LX:⍝ EXIT THE LOOP
⍝ UPDATE PARTITION VECTOR COUNT
 COUNT_PARTITIONS 6 1
⍝ LOAD PARTITION VECTOR IN MAIN MEMORY
 5 SETMEMORY PVCT
 SHOWMACHINE ⍝ DISPLAY MACHINE'S CONTENTS
```

CALC_WITH_PARTITIONS function applies a specified function to each partition.

```
 CALC_WITH_PARTITIONS AR;⎕IO;S;P;LN;LP;TP;TLN;L;V;D
ᴀ DESCRIPTION: APPLIES A SPECIFIED FUNCTION TO EACH PARTITION
 ⎕IO←1 ᴀ ORIGIN
ᴀ INPUT: 2 COLUMNS ARRAY AR(COLUMN1=GROUP BY DATA (BINS); COLUMN2= DATA TO PARTITION)

ᴀ PARSE OUT AR INTO 2 VARIABLES:
 V←AR[;1] ᴀ GROUP BY (BINS) VECTOR
 D←AR[;2] ᴀ DATA TO PARTITION
 26 SETMEMORY V     ᴀ LOAD V IN MAIN MEMORY AT ADDRESS 26
 51 SETMEMORY ¯1⌽V ᴀ LOAD ROTATED V IN MAIN MEMORY AT ADDRESS 51
ᴀ ROTATED V HAS THE LAST ELEMENT OF V AT THE FIRST POSITION
 L←⍴V    ᴀ FIND THE LENGTH OF V
 2 SETMEMORY L ᴀ LOAD L IN MAIN MEMORY AT ADDRESS 2
 3 SETMEMORY 26ᴀ V MAIN MEMORY ADDRESS
 4 SETMEMORY 51ᴀ ¯1⌽V MAIN MEMORY ADDRESS
 5 SETMEMORY L ᴀ PARTITION COUNT ADDRESS IN MAIN MEMORY
 6 SETMEMORY 76ᴀ PARTITION VECTOR ADDRESS IN MAIN MEMORY
 LOADPVCT 5 ᴀ LOAD PARTITION VECTOR COUNT
LOOP:LOADVCT 2 ᴀ LOAD VECTOR COUNT
 ->(0=VCT)↑LX ᴀ EXIT IF VECTOR COUNT IS ZERO
 LOADV 1 3 1 ᴀ LOAD V1 VECTOR REGISTER WITH V
 LOADV 2 4 1 ᴀ LOAD V2 VECTOR REGISTER WITH ¯1⌽V
 1(F3 DJBOX)0 ᴀ USE DJBOX OPERATOR TO CALCULATE PARTITION VECTOR FOR V
 STOREV 3 6 1 ᴀ STORE PARTITION VECTOR IN MAIN MEMORY
 ->LOOP ᴀ LOOP BY VECTOR COUNT
LX:ᴀ EXIT THE LOOP

ᴀ UPDATE PARTITION VECTOR COUNT
 COUNT_PARTITIONS 6 1
ᴀ LOAD PARTITION VECTOR IN MAIN MEMORY
 5 SETMEMORY PVCT
ᴀ LOAD D(DATA TO PARTITION) IN MAIN MEMORY
 26 SETMEMORY D
 2 SETMEMORY L ᴀ LOAD L IN MAIN MEMORY
 3 SETMEMORY 26 ᴀ D MAIN MEMORY ADDRESS
 4 SETMEMORY 76 ᴀ PARTITION VECTOR ADDRESS IN MAIN MEMORY
 7 SETMEMORY 51 ᴀ RESULT ADDRESS IN MAIN MEMORY OF APPLYING A SPECIFIED FUNCTION TO
ᴀ EACH PARTITION
ᴀ PARTITION VECTOR DATA USING GIVEN PARTITION VECTOR
 (P LN)←CALC_PARTITIONS 2 3 4 1
 LP←0 ᴀ LOOP COUNTER
LOOP2:LOADLPVCT 5 ᴀ LOAD PARTITION VECTOR COUNT
 ->(0=PVCT)↑LX2 ᴀ EXIT IF PARTITION VECTOR COUNT IS ZERO
ᴀ RETURN THE MOST PSS (PARTITION SECTION SIZE) PARTITIONS IN TP
ᴀ WITH RESPECTIVE PARTITION LENGTHS IN TLN
 (TP TLN)←PREPARE_T_PARTITIONS(LP P LN)
ᴀ LOAD Deutsch-JOZSA VECTOR REGISTERS
 TP LOAD_PARTITIONS_INTO_DJVR TLN
ᴀ APPLY FUNCTION F2 TO EACH PARTITION USING DJBOX OPERATOR
 3(F2 DJBOX)1
 PSTOREV 3 7 1 ᴀ STORE RESULT OF APPLYING F2 TO PARTITIONS IN MAIN MEMORY
 LP←LP+1 ᴀ INCREMENT LOOP COUNTER
 ->LOOP2 ᴀ LOOP BY PARTITION VECTOR COUNT
LX2:ᴀ EXIT THE LOOP
 SHOWMACHINE ᴀ DISPLAY MACHINE'S CONTENTS
```

DJBOX operator simulates essential elements of the quantum circuit for the Deutsch-Jozsa algorithm.

```
A(F DJBOX)B;⎕IO;STATES;P;X;Y;T;TT;CONV;Z;TTT;K;I;P3CONV
⍝ DESCRIPTION: SIMULATE ESSENTIAL ELEMENTS OF THE QUANTUM CIRCUIT FOR
⍝ THE Deutsch-JOZSA ALGORITHM
 ⎕IO←1 ⍝ ORIGIN
⍝ INPUT: A IS VECTOR REGISTER
⍝        B IS VECTOR REGISTER OR Deutsch-JOZSA VECTOR REGISTER
⍝        F IS FUNCTION USED TO CALCULATE PARTITION VECTOR
⍝        OR TO APPLY TO EACH PARTITION
 STATES←∊HNGATE QR⍴0 ⍝ PREPARE ALL THE QUERY REGISTER QUBITS IN THE |0> STATE
⍝ AND APPLY HADAMARD TRANSFORM ON THE QUERY REGISTER
⍝ THE QUERY REGISTER WILL BE A SUPERPOSITION OF ALL VALUES
 P←(STATES='+')∨(STATES='-') ⍝ IDENTIFY WHERE IS + OR -
 X←⍕¨(QRSIZE QR)⍴(~P)/STATES ⍝ CREATE AN ARRAY OF ALL SUPERPOSITION VALUES IN
⍝ THE QUERY REGISTER
 Y←HNGATE 1 ⍝ APPLY HADAMARD GATE ON THE ANSWER REGISTER
 P3CONV←(0 QRSIZE)⍴0 ⍝ INITIALIZE CONTROL VARIABLE THAT WILL BE USED TO
⍝ DETERMINE IF FUNCTION F IS BALANCED OR CONSTANT
 R1←A ⍝ SET GLOBAL VARIABLE FOR THE VECTOR REGISTER A
 R0←B ⍝ SET GLOBAL VARIABLE FOR THE REGISTER B
 I←0  ⍝ INITIALIZE OUTER LOOP VARIABLE
L0:->(QRSIZE<I←I+1)↑LX ⍝ LOOP BY THE QUERY REGISTER STATES
 LOAD_QUERY_REGISTER X[I;] ⍝ LOAD THE QUERY REGISTER WITH STATE X[I;]
⍝ NOTE: X IS AN ARRAY OF ALL SUPERPOSITION VALUES IN THE QUERY REGISTER
 T←HNGATE X[I;] ⍝ APPLY HADAMARD GATE ON THE STATE X[I;]
 P←~∊(T='+')∨(T='-') ⍝ IDENTIFY WHERE IS NOT + OR - (FIND STATES)
 TT←P⊂T ⍝ ENCLOSE STATES
 CONV←2⊥⍉⍕¨⊃TT ⍝ EXPRESS EACH STATE IN DECIMAL REPRESENTATION (STATES ARE IN BINARY
REPRESENTATION)
 Z←(~P)/∊T ⍝ COLLECT ALL +,-
 K←0 ⍝ INITIALIZE INNER LOOP VARIABLE
 TTT←'' ⍝ INITIALIZE VARIABLE WHICH HOLDS RESULT OF LOOP ITERATION
L1:->((QRSIZE-1)<K←K+1)↑L2 ⍝ LOOP QRSIZE-1 TIMES, BEGIN LOOP WITH THE SECOND STATE
 TTT←TTT,(1+Z[K]='+')⊃(¯1×CONV[1+K])(CONV[1+K])⍝ IF + SIGN ADD STATE,
⍝ IF - SIGN, ADD STATE MULTIPLIED BY ¯1
 ->L1 ⍝ GO TO NEXT LOOP ITERATION
L2: ⍝ EXIT LOOP
 P3CONV←P3CONV,[1](Y YPLUSF(F X[I;]))×(0,TTT) ⍝ ADD 0 STATE TO TTT,
⍝ MULTIPLY STATES WITH THE OUTPUT OF Y ADD MODULO 2 F(X) AND
⍝ ADD THEM TO P3CONV
 ->L0 ⍝ GO TO NEXT STATE IN THE QUERY REGISTER
LX:⍝ EXIT OUTER LOOP
 T←+⌿P3CONV ⍝ SUM VALUES IN EACH COLUMN OF P3CONV
 ->(QRSIZE=(+/T=0))↓L3 ⍝ CHECK IF T HAS ALL ZERO'S
⍝ IF NOT JUMP TO LABEL L3
 'QUERY REQISTER IN 0 STATE'
 LOAD_QUERY_REGISTER 0 0 0 ⍝ LOAD QUERY REGISTER WITH ZERO STATES
 ->L4 ⍝ EXIT
L3: ⍝ T DOES NOT HAVE ALL ZERO'S
 TT←(T≠0)/(0,⍳(QRSIZE-1)) ⍝ IDENTIFY WHICH STATES ARE IN SUPERPOSITION IN THE QUERY
REGISTER
⍝ AND EXPRESS EACH OF THESE STATES IN BINARY REPRESENTATION
 'QUERY REGISTER EVALUATED IN SUPERPOSITION OF THESE STATES ',⊂CONVERT_TO_BINARY TT
 LOAD_QUERY_REGISTER QR⍴'⊛' ⍝ DISPLAY SUPERPOSITION WITH ⊛
L4: ⍝ EXIT
```

YPLUS function adds modulo 2 a state and function f(x).

```
R←Y YPLUSF F;⎕IO;P;V
⍝ DESCRIPTION:ADD MODULO 2 STATE Y AND FUNCTION F
 ⎕IO←1 ⍝ ORIGIN
⍝ INPUT: STATE Y AND FUNCTION F

 P←~∊(Y='+')∨(Y='-') ⍝ IDENTIFY WHERE IS NOT + OR - (FIND STATES)
 V←2|(⍕¨(∊P)/∊Y)+F ⍝ ADD F TO EVERY STATE OF Y, DIVIDE BY 2
⍝ AND DETERMINE REMINDER FOR EVERY STATE (ADD MODULO 2)
 R←(1+V[1]=1)⊃1 ¯1 ⍝ RETURN 1 IF THE FIRST COMPONENT OF V IS 0,
⍝ RETURN ¯1 IF THE FIRST COMPONENT OF V IS 1
```

F3 function compares two vector register cell contents and store the result in the third vector register cell.

```
 R←F3 B;⎕IO;A
⍝ DESCRIPTION: COMPARE CONTENTS OF TWO VECTOR REGISTER CELLS
⍝ AND STORE COMPARISON RESULT (0 OR 1) IN THE THIRD VECTOR
⍝ REGISTER CELL
 ⎕IO←0 ⍝ ORIGIN

⍝ INPUT: B - QUERY REGISTER STATE IN BINARY REPRESENTATION
⍝ CONVERT BINARY INPUT B INTO ADDRESS A
 A←2⊥B
 →(VR[R0;A]='.')↑L0 ⍝ IF CELL AT ADDRESS A FOR VECTOR REGISTER R0 IS EMPTY
⍝ JUMP TO L0, NOTHING TO COMPUTE
⍝ COMPARE R0 VECTOR REGISTER CELL AT THE ADDRESS A
⍝ WITH R1 VECTOR REGISTER CELL AT THE ADDRESS A
⍝ IF CONTENTS ARE THE SAME PICK 0
⍝ AND STORE 0 IN THE THIRD VECTOR REGISTER CELL
⍝ AT THE ADDRESS A
⍝ OTHERWISE STORE 1
 VR[2;A]←(VR[R0;A]≠VR[R1;A])⊃0 1
L0:
⍝ RETURN ALWAYS 1
 R←1
```

F2 function sums elements of a vector stored in a Deutsch-Jozsa vector register and stores the sum in a vector register.

```
 R←F2 B;⎕IO;A;P;T
⍝ DESCRIPTION: SUM VECTOR ELEMENTS STORED IN
⍝ Deutsch-JOZSA VECTOR REGISTER 1+A AND STORE SUM
⍝ IN VECTOR REGISTER R1 AT ADDRESS 1+A
 ⎕IO←1 ⍝ ORIGIN

⍝ INPUT: B - QUERY REGISTER STATE IN BINARY REPRESENTATION
⍝ CONVERT BINARY INPUT B INTO ADDRESS A
 A←2⊥B
 →(DJVR[1+A;R0]='.')↑L0 ⍝ IF CELL AT ADDRESS R0 FOR
⍝ Deutsch-JOZSA VECTOR REGISTER 1+A IS EMPTY JUMP TO L0, NOTHING TO COMPUTE
 P←DJVR[1+A;R0] ⍝ READ OFF HOW MANY ELEMENTS ARE IN VECTOR (CELL R0)
 VR[R1;1+A]←+/DJVR[1+A;1+⍳P] ⍝ SUM THOSE ELEMENTS AND STORE RESULT INTO
⍝ VECTOR REGISTER R1 AT ADDRESS 1+A
L0:
⍝ RETURN ALWAYS 1
 R←1
```

CLEARMEMORY function clears main memory between two addresses.

```
 R←S CLEARMEMORY E;⎕IO
⍝ DESCRIPTION: CLEAR MAIN MEMORY BETWEEN ADDRESSES S AND E, INCLUDING S AND E
 ⎕IO←1 ⍝ ORIGIN
⍝ INPUT: MAIN MEMORY ADDRESSES S AND E

 ((1+E-S)↑(S-1)↓MM)←(1+E-S)⍴'.' ⍝ CLEAR MAIN MEMORY
```

CONVERT_TO_BINARY function creates the binary representation of a number in the decimal representation. It is basically "Convert integer to binary" APL2 idiom from the APL2 Idiom Library imbedded into a function.

```
 R←CONVERT_TO_BINARY N;⎕IO
⍝ DESCRIPTION: CREATE BINARY REPRESENTATION OF THE NUMBER
⍝ IN DECIMAL REPRESENTATION
 ⎕IO←1 ⍝ ORIGIN
⍝ INPUT: NUMBER N IN DECIMAL REPRESENTATION

 R←⌽((1+⌊2 ⍟ 1⌈⌈/N)⍴2)⊤N ⍝ CREATE BINARY REPRESENTATION FOR N
```